\documentclass[fleqn,twoside]{article}
\usepackage{espcrc2}

\usepackage{graphicx}
\newcommand{\AmS}{{\protect\the\textfont2
  A\kern-.1667em\lower.5ex\hbox{M}\kern-.125emS}}

\title{The X-ray emission of the Crab-like pulsar PSR J0537$-$6910}

\author{T. Mineo\address[MCSD]{Istituto di Astrofisica Spaziale e Fisica
       Cosmica, CNR, Sezione di Palermo  \\
       via U. La Malfa 173, I-90139, Palermo, Italy},
        G. Cusumano\addressmark
        and
        E. Massaro\address[MCSG]{Dipartimento di Fisica, Universit\'{a}
       la Sapienza, \\
       Piazzale A. Moro 2, I-00185, Roma, Italy}}

\begin{document}

\begin{abstract}
In this paper we present some preliminary result on the spectral and timing analysis of
the X-ray pulsed emission from the 16 ms pulsar PSR J0537$-$6910 in the energy range 0.1--30
keV, based on archival BeppoSAX and RossiXTE observations. This pulsar,
discovered by
Marshall et al. \cite{mar98} in the LMC field with RXTE, is the fastest spinning pulsar
associated with a supernova remnant. It is characterized by strong glitch activity with
the highest rate of all known Crab-like system.
\vspace{1pc}
\end{abstract}

% typeset front matter (including abstract)
\maketitle

\section{Introduction}

PSR J0537$-$6910 is a Crab-like pulsar embedded in the supernova remnant
N157B in the 30 Doradus region
located at the centre of the LMC. This supernova belongs to the
rare class of Crab-like remnants \cite{sew89} distinguished by
their centrally-filled morphology and non-thermal X-ray spectra,
whose emission is predominantly  powered by the young
pulsar at the center.
PSR J0537$-$6910 is the most rapidly rotating pulsar associated
with a SNR ($P$=16.1 ms), twice as fast as the Crab pulsar,
with a period derivative of $\dot{P}$=5.17$\times$10$^{-14}$ s s$^{-1}$,
a rotational energy loss rate of $\dot{E}=4.8\times10^{38}$ erg s$^{-1}$,
a magnetic field at the neutron star surface of $B_s=9.2\times10^{11}$ G and
a braking age of
$\tau=5.0\times10^{3}$ yr in agreement with the predicted SNR age \cite{wan98a}.
It represents the oldest known Crab-like pulsar.

Pulsed emission was first discovered in the X-ray range by RXTE/PCA
\cite{mar98} and soon after confirmed by BeppoSAX \cite{cus98}
and ROSAT/HRI \cite{wan98b}. The light curve is characterized by a
narrow pulse shape with a full width duty cycle of about 10\%
\cite{wan01,wan98b}.
The 2--10 keV pulsed spectrum has been modeled with a simple power law:
RXTE+ASCA measured a spectral index of 1.6$\pm$0.4 and a 2--10 flux of
(6.7$\pm$0.6)$\times$10$^{-13}$ erg cm$^{-2}$
 s$^{-1}$  \cite{mar98}, in agreement with the
BeppoSAX  values of 1.1$\pm$0.4 and
(5.1$\pm$1.1)$\times$10$^{-13}$ erg cm$^{-2}$ s$^{-1}$
\cite{cus98}.

The source shows intense glitch activity: 6 glitch events, with a mean amplitude in the
frequency change of $\Delta \nu/\nu $=0.36$\times$10$^{-6}$,
have been observed during the almost 3 years monitoring
campaign performed by RXTE \cite{got02}.
No pulsed emission have been detected at radio wavelengths since
now \cite{wan01}.

\section{Observation and Data Reduction}
The RXTE observations considered in our analysis were performed
between 26 March and 19 December 2001.
We used  PCA \cite{jah94}
accumulated in "Good Xenon"  telemetry mode (1$\mu$s accuracy)
selecting all detector layer to increase the S/N ratio of the pulsation.
Standard selection criteria were applied on the observation data
excluding time intervals correspondent to South Atlantic Anomaly
passage and when Earth limb was less than 10$^{o}$ and angular
distance between the source position and the pointing of the
satellite was larger than 0.02$^{o}$.
All  analysed
observations were pointing to PSR J0537$-$6910 and are relative to the same
gain epoque. The total exposure is about 1300 ks.

The narrow field instruments onboard BeppoSAX \cite{boe97}
observed LMC field several times during its life. For our analysis, we
selected the observations
pointing at N157B and considered data from the two imaging detectors LECS (0.1--10 keV)
and MECS (1.6--10 keV) for a total exposure of 29 ks and 89 ks, respectively.  In both
instruments, events were extracted from a circular region centred at the source position with
a radius of 3' which maximize the signal-to-noise ratio of the pulsed component.
The LECS energy range 0.1--1.0 keV was not included in the
analysis because of the low statistical significance of the pulsed
emission due to the presence of a contribution from the
hot gas of 30 Doradus \cite{cus98}.

\section{Timing Analysis}
The UTC arrivals times of all selected events were first converted
to the Solar System Barycentre using the (J2000) pulsar position
$\alpha$=5$^h$ 37$^m$ 47.2$^s$ and $\delta$=-69$^o$ 10' 23"
\cite{wan98b} and the JPL2000 planetary ephemeris (DE200, \cite{sta82}).
For each observation, we searched the pulsed
frequency $\nu$ using the folding  technique in a range centred at
the value computed with the pulsar ephemeris reported by
\cite{got02}. The central time of each observation was chosen as
reference epoque and the corresponding frequency was estimated by
fitting the $\chi^2$ peak with a gaussian. Frequency errors at
1$\sigma$ level were computed from the interval corresponding to a
unit decrement with respect to the maximum in the $\chi^2$ curve.
The top panel of Fig. 1  shows the frequencies  obtained for BeppoSAX (triangles)  and
RXTE (stars) together with the linear ephemeris quoted by
\cite{got02} (dotted line);
residuals respect to this relation are presented in the bottom panel.

The highest significance of
the pulsation is reached in the energy intervals
2.5--30 keV in RXTE data and 1.0--10 keV in BeppoSAX data.
The RXTE  profile is shown in Fig. 2: it
is characterized by a narrow single peak with a duty cycle of 0.28
at zero level. Fitting the peak with a gaussian,  a value of
0.11$\pm$0.07 is obtained for the FWHM, compatible with the value
obtained with ROSAT \cite{wan98b} and Chandra data \cite{wan01}

\begin{figure}[htb]
\includegraphics[angle=-90,width=18pc]{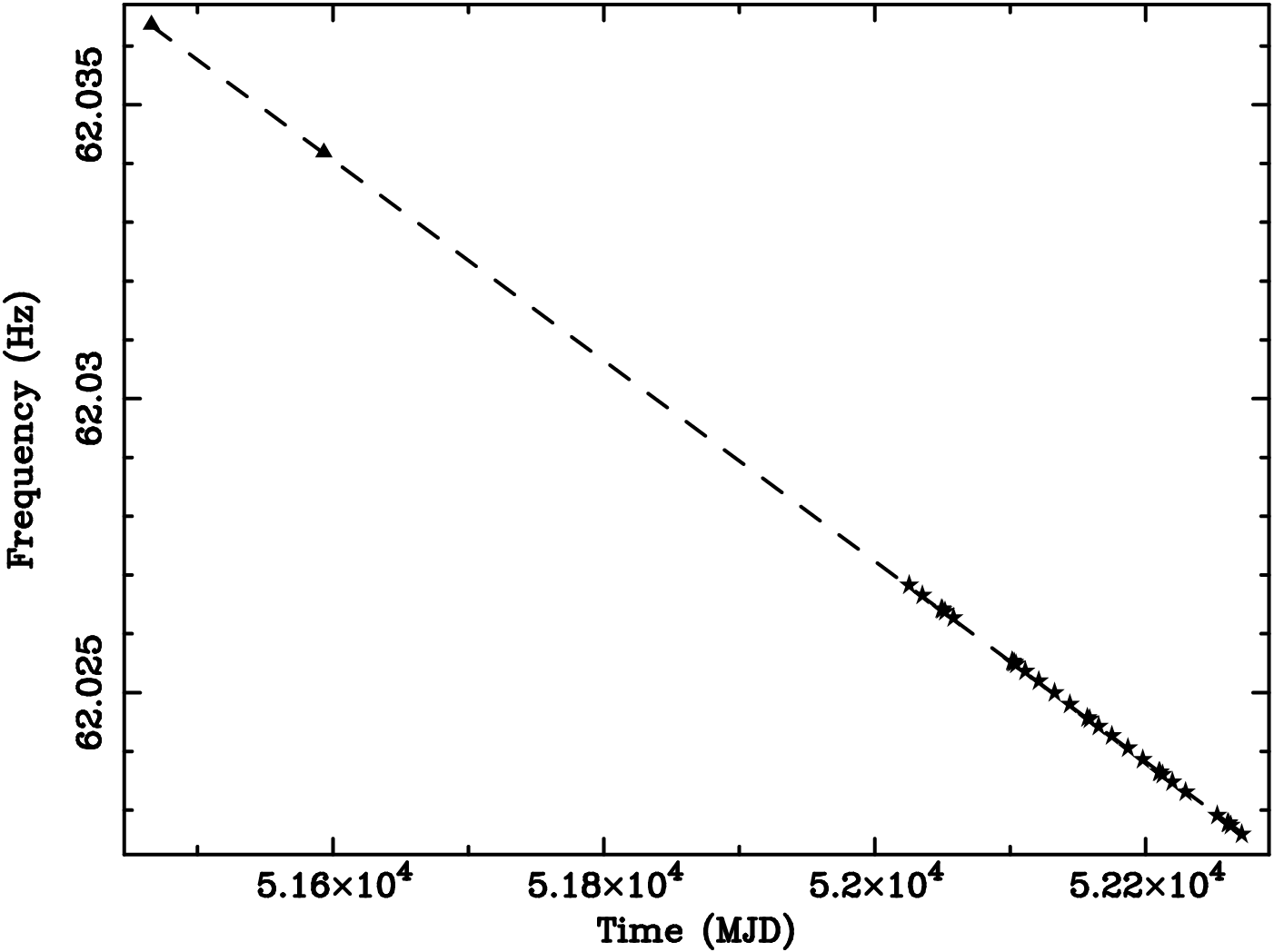}
\includegraphics[angle=-90,width=18pc]{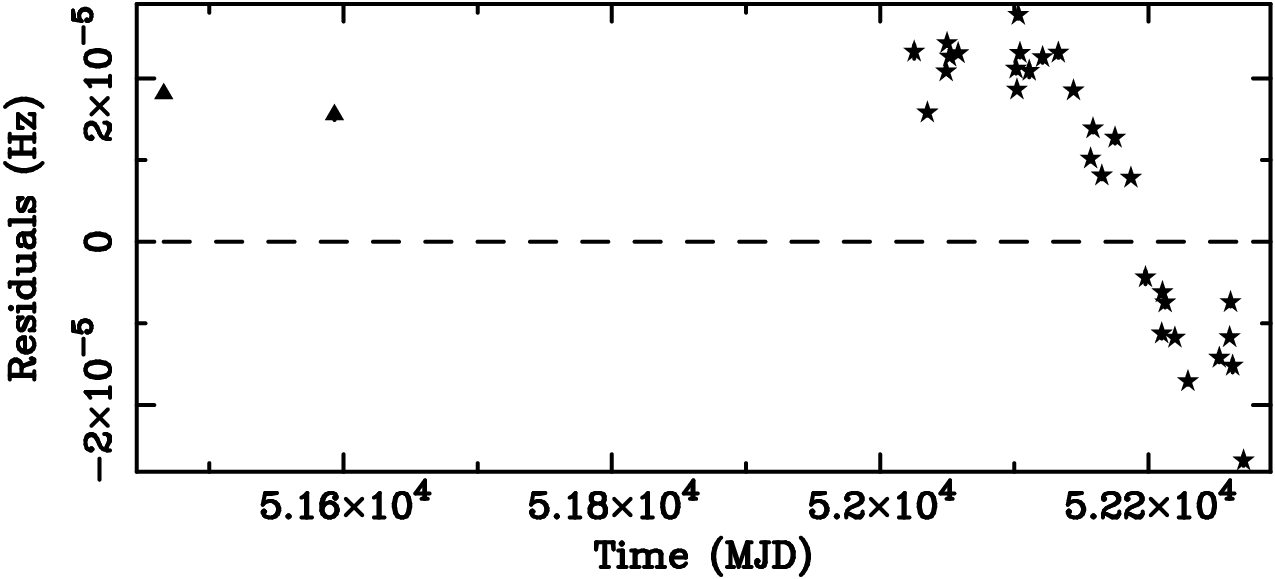}
\caption
{PSR J0537$-$6910 frequency evolution (top panel) and  residuals
(bottom panel) respect to the \cite{got02} relation (dotted line). BeppoSAX data are
indicated with triangles,  RXTE data with stars}
\label{fig1}
\end{figure}

\section{Spectral analysis}
Pulse phase histograms were accumulated for each unit and for each observation of the PCA
independently for the 256 PHA channels. The same procedure has
been applied to MECS data.  RXTE response matrix were derived for
each unit and summed together weighting with the background
subtracted counts of the correspondent PCU's phase histograms.
RXTE spectra were combined by summing the individuals units and
assigning a total exposure time equal to the sum of the
individual exposures.
The spectral distribution of the pulsed signal was accumulated in
the  phase interval 0.64--0.84 subtracting the off-pulse
level evaluated in the phase range 0.0--0.5.
The two phase regions considered in our analysis are indicated in Fig. 2.

\begin{table*}[htb]
\caption{Best fit parameters}
\label{table1}
\newcommand{\cc}[1]{\multicolumn{1}{c}{#1}}
\renewcommand{\tabcolsep}{2pc} % enlarge column spacing
\renewcommand{\arraystretch}{1.2} % enlarge line spacing
\begin{tabular}{lcc}
\hline
                                 & Simple power law & Curved power law \\
\hline
$N_{\rm H}$ (cm$^{-2}$)          & (1.3$\pm$0.5)$\times$10$^{22}$ & (1.4$\pm$0.7)$\times$10$^{22}$ \\
$\alpha$                         & 1.67$\pm$0.07 & 1.33$\pm$0.12 \\
$\beta$                          & --            &  0.15$\pm$0.05           \\
$\chi^2$ (dof)                   &  67.2 (82) &  60.2 (81)\\
\hline
\end{tabular}
%\\[2pt]
%The experimental values are given in ref. \cite{Eato75}.
\end{table*}

The BeppoSAX and RXTE spectra covering the
energy range 1.0--30 keV were fitted together. The
intercalibration factors were fixed to the values derived fitting
the common energy interval 2-10 keV:  in particular the value
0.9$\pm$0.3 has been derived for LECS-MECS and 1.19$\pm$0.09 for
XTE-MECS. The spectra have been modeled with a simple power law
whose best fit parameters are shown in Table 1.
The best fit parameters of the simple power law are in agreement
with the previous evaluations \cite{cus98,wan98b} and, according to
the $\chi^2$ value, this model well represent the spectra
over the whole range. The detected 2--10 unabsorbed flux is
(7.3$\pm$0.5)$\times$10$^{-13}$ erg cm$^{-2}$ s$^{-1}$, corresponding to a
uniform luminosity of 1.9$\times$10$^{35}$ erg  s$^{-1}$
for an estimate distance of 47 kpc \cite{gou95}. Table 1 shows the
best fit result.
A second models have been investigated: a
 curved power law characterized by a linear dependence of
the spectral slope upon the logarithm of energy:

\begin{equation}
 F(E)=K\,E^{-[\alpha \,+\,\beta \, Log(E)]}
\end{equation}

\begin{figure}[htb]
\includegraphics[angle=-90,width=18pc]{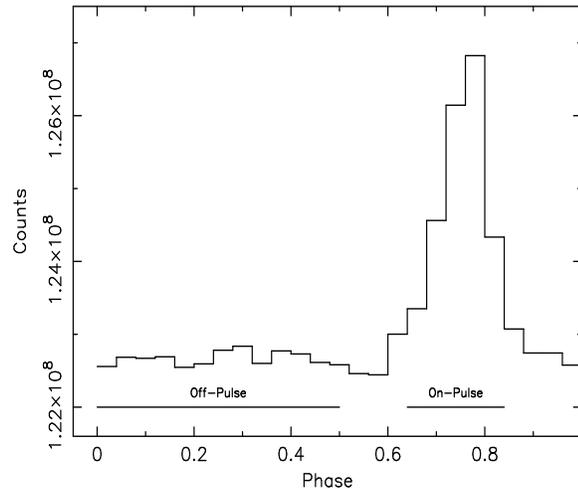}
\caption{PSR J0537$-$6910  pulse profile in the energy range 2.5--30 keV
obtained with RXTE data.
The on-sourse and background phase intervals used for the
spectral analysis are also indicated.}
\label{fig2}
\end{figure}
\noindent
This model has been used to fit the broad band pulsed spectra of the
other Crab-like pulsars Crab \cite{mas00}, PSR B1509-58\cite{cus01}
PSR B0540-69\cite{dep03}.
The best fit result for this model is also shown in Table 1.
The introduction of the curvature
parameter $\beta$ however is significant at 99.7\% confidence level computed
with the F-test for the introduction of a single parameter.
Moreover the $\beta$ detected value is in agreement with the
values obtained for the three other pulsars.

\section{Conclusion}
We presented some preliminary results on the spectral and timing analysis of the X-ray
pulsed emission of the 16 ms pulsar PSR J0537$-$6910 in the energy range 1.0--30 keV, based on
archival BeppoSAX and RossiXTE observations.
The  detected profile
is characterized by a narrow single peak with a duty cycle of 0.28
at zero level.
The pulsed spectrum has been fitted with two models: the simple power law and a curved
power law. The simple power law gave acceptable $\chi^2$, but the introduction of the curvature
parameter $\beta$ is significant at 99.7\% confidence level.
This result suggests that the photon index softens at higher
energies.
The value of the bending parameters is compatible within the
errors with the values find for the other Crab-like pulsars Crab \cite{mas00},
PSR B1509-58 \cite{cus01}, PSR B0540-69 \cite{dep03}.
According to the measured parameters of the curved model, the SED of the pulsed emission
have a maximum at the energy:

\begin{equation}
E_m=10^{[(2\,-\,\alpha)/(2\,\beta)]}
\end{equation}

\noindent
wich correspond to $\sim$170 keV,  practically coincident with the value measured
in the interpulse of the Crab pulsar \cite{cus01}.

The models that have been proposed to explain high energy emission
from pulsars can be classified in two distinct classes, outer gap
\cite{che86} and
polar gap models \cite{dau94}, depending
on the location of the emitting region. In the framework of the
outer gap model, the energy of the maximum follows the appropriate
scaling relation with respect to the interpeak emission of the Crab
 as observed for the PSR B1509-58 \cite{cus01}.

\end{document}